\documentclass[10pt, aps, prd, twocolumn, superscriptaddress, nofootinbib]{revtex4-2}

\usepackage{amssymb,amsmath,mathrsfs,enumerate}
\usepackage{graphicx}
\usepackage{feynmp-auto}
\usepackage{mathtools}
\usepackage{caption}
\usepackage{subcaption}
\usepackage[dvipsnames]{xcolor}
\usepackage[colorlinks=true, linkcolor=blue, urlcolor=Blue, citecolor=Mulberry]{hyperref}
\usepackage{orcidlink}
\usepackage{multirow}
\usepackage{booktabs}
\usepackage[normalem]{ulem}
\usepackage{microtype}  
\usepackage{ragged2e}   
\usepackage{bm}

\usepackage{tikz}
\usetikzlibrary{arrows.meta, positioning, shapes.geometric}

\makeatletter
\patchcmd{\@makecaption}{\ignorespaces}{\justifying\ignorespaces}{}{}
\makeatother

\begin{document}

\title{Probing Axion Dark Matter via the Chiral Magnetic Effect in Zero-Bias Weyl Semimetals}

\author{Debajit Bose \orcidlink{0000-0001-8594-8885}}
\email{debajitbose550@gmail.com}
\affiliation{Centre for High Energy Physics, Indian Institute of Science, C. V. Raman Avenue, Bengaluru 560012, India}

\author{Prataya Chandra \orcidlink{0009-0004-1819-7769}}
\email{prataya2000@gmail.com}
\affiliation{Department of Physics, Indian Institute of Technology Kharagpur, Kharagpur 721302, India}

\author{Sudhansu S. Mandal \orcidlink{0000-0003-0665-0218}}
\email{sudhansu@phy.iitkgp.ac.in}
\affiliation{Department of Physics, Indian Institute of Technology Kharagpur, Kharagpur 721302, India}

\author{Tirtha Sankar Ray \orcidlink{0000-0002-2085-7429}}
\email{tirthasankar.ray@gmail.com}
\affiliation{Department of Physics, Indian Institute of Technology Kharagpur, Kharagpur 721302, India}
\date{\today}
%
%
\begin{abstract}

Sub-eV axion dark matter behaves as a coherent classical field that can induce macroscopic current in quantum materials. We explore the possibility of axion detection via the chiral magnetic effect in zero-bias Weyl semimetals under a static external magnetic field. We demonstrate that for a $1 \, {\rm cm^2}$   sample in a realistic $10 \, {\rm T}$ magnetic field, the signal remains in the observable femto-ampere range. Utilizing state-of-the-art SQUID-based current readout, the setup can probe axion--electron couplings below existing stellar cooling bounds across a broad range of axion masses.

\end{abstract}
%
%
\maketitle
%
%

%
%
\section{Introduction}
\label{sec:intro}

A wide range of astrophysical and cosmological observations strongly support the existence of dark matter (DM) in the Universe\,\cite{Planck:2018vyg, Bertone:2004pz}. Detection of weak scale dark matter relies on the measurement of transfer of momentum imparted on target detector material due to a DM wind and at present is constrained by the null results of direct detection experiments\,\cite{XENON:2025vwd, LZ:2024zvo}. However, conventional direct detection experiments lose sensitivity to light DM particles due to the small momentum transfer, often below the experimental threshold. Consequently, searches for light DM are better suited to probe energy shifts induced by the coherent oscillating background DM field. Among the motivated light DM candidates are  the axions that were originally  introduced in the context  of the strong CP problem\,\cite{Peccei:1977hh, Weinberg:1977ma, Wilczek:1977pj}, consequently  many beyond Standard Model scenarios predict similar light scalar  particles\,\cite{Svrcek:2006yi, Conlon:2006tq, Arvanitaki:2009fg}, collectively known as axion-like particles (ALPs)\footnote{Hereafter, we use the terms ``axion" and ``ALP" interchangeably.}. ALPs can account for the observed DM relic abundance through the misalignment mechanism of a coherent field\,\cite{Preskill:1982cy, Abbott:1982af, Co:2019jts}.

The experimental landscape for ALPs has seen extensive activity in recent years\,\cite{Marsh:2015xka, Graham:2015ouw, Irastorza:2018dyq, Irastorza:2021tdu, Berlin:2023ubt}. An increasingly interdisciplinary frontier of tabletop experiments has emerged, leveraging advances in quantum-limited measurement and precision optics to probe ALP-like coherent DM fields. These include searches via atomic and molecular transitions\,\cite{Sikivie:2014lha, Santamaria:2015gro, Braggio:2017oyt}, resonant LC circuits\,\cite{Sikivie:2013laa, Kahn:2016aff}, nuclear magnetic resonance spin-precession\,\cite{Graham:2011qk, Budker:2013hfa, Arvanitaki:2014dfa}, bulk effects in quantum materials and emerging condensed matter systems\,\cite{Safronova:2017xcl, Berlin:2024pzi}, Casimir effect and short-range force measurements\,\cite{Mostepanenko:2020lqe, Klimchitskaya:2021gkd}, and precision polarimetry experiments\,\cite{DellaValle:2015xxa, DeRocco:2018jwe}.

%
%
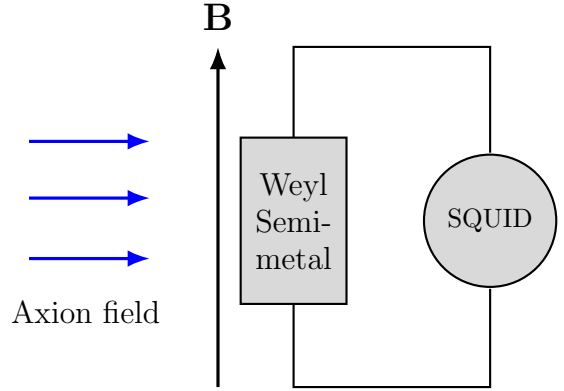
\begin{figure}[ht]
\centering

\begin{tikzpicture}[
    >=Latex,
    thick,
    scale=0.5,
    every node/.style={font=\Tiny}
]

\draw[blue, very thick, ->] (-5,1.5) -- (-1.8,1.5);
\draw[blue, very thick, ->] (-5,0.0) -- (-1.8,0.0);
\draw[blue, very thick, ->] (-5,-1.6) -- (-1.8,-1.6);

\node[font=\fontsize{28}{30}\large] at (-3.5,-3.0)
{Axion field};

\draw[very thick, ->] (0,-5) -- (0,4);
\node[font=\Large] at (-0.0,4.8) {$\mathbf{B}$};

\draw[fill=gray!30, thick]
(0.6,-2.8) rectangle (3.4,1.6);

\node[align=center,font=\large]
at (2.0,-0.6)
{Weyl\\Semi-\\metal};

\draw[thick]
(2.0,1.6) -- (2.0,4.0)
-- (7.2,4.0)
-- (7.2,1.2);

\draw[thick]
(2.0,-2.8) -- (2.0,-5.0)
-- (7.2,-5.0)
-- (7.2,-2.4);

\draw[fill=gray!30, thick] (7.2,-0.6) circle (1.75);

\node[font=\normalsize] at (7.2,-0.6) {SQUID};

\end{tikzpicture}

\caption{\justifying Schematic of the proposed setup. The axion field induces a signal in the Weyl semimetal placed in an external magnetic field $\mathbf{B}$, which is read out by a SQUID.}
\label{fig:setup}
\end{figure}
%
%

In this article, we explore the possibility of searching for sub-eV axion DM\,\cite{Peccei:1977hh, Weinberg:1977ma, Wilczek:1977pj, Sikivie:2006ni, Raffelt:2006cw} with chiral magnetic effect (CME) current in Weyl semimetals due to axion-electron coupling\,\cite{Zhitnitsky:1980tq, Dine:1981rt, Kim:1979if, Shifman:1979if, Choi:2021kuy}. Electrons in the momentum-space separated topologically protected opposite chiral nodes of the Weyl semimetals, experience oppositely shifted chemical potential induced by the axion background. This leads to a bulk CME current in the presence of an external magnetic field $\mathbf{B}$. We demonstrate that axion driven CME current density is given by\,\cite{Fukushima:2008xe, Armitage:2017cjs, Chen:2013iga}
\begin{equation}\label{eq:J_CME}
    \mathbf{J}_{\rm CME} = \frac{e^2}{2\pi^2}\,\mu_5 \, \mathbf{B} \, .
\end{equation}
This phenomenon may be exploited in zero-bias experiments\,\cite{Kharzeev:2013ffa} as schematically illustrated in Fig.~\ref{fig:setup}, and complements axion driven CME proposals in conventional materials \cite{Hong:2022nss, Hong:2025raa}. Weyl semimetals in the presence of an external magnetic field act as axionic transducers, and the observation of  such a  zero-bias  CME current can probe previously unexplored axion parameter space, as shown in Fig.~\ref{fig:gae_vs_B}.
%
%
%

%
%
\begin{figure}[h]
    \centering
    \includegraphics[width=\linewidth]{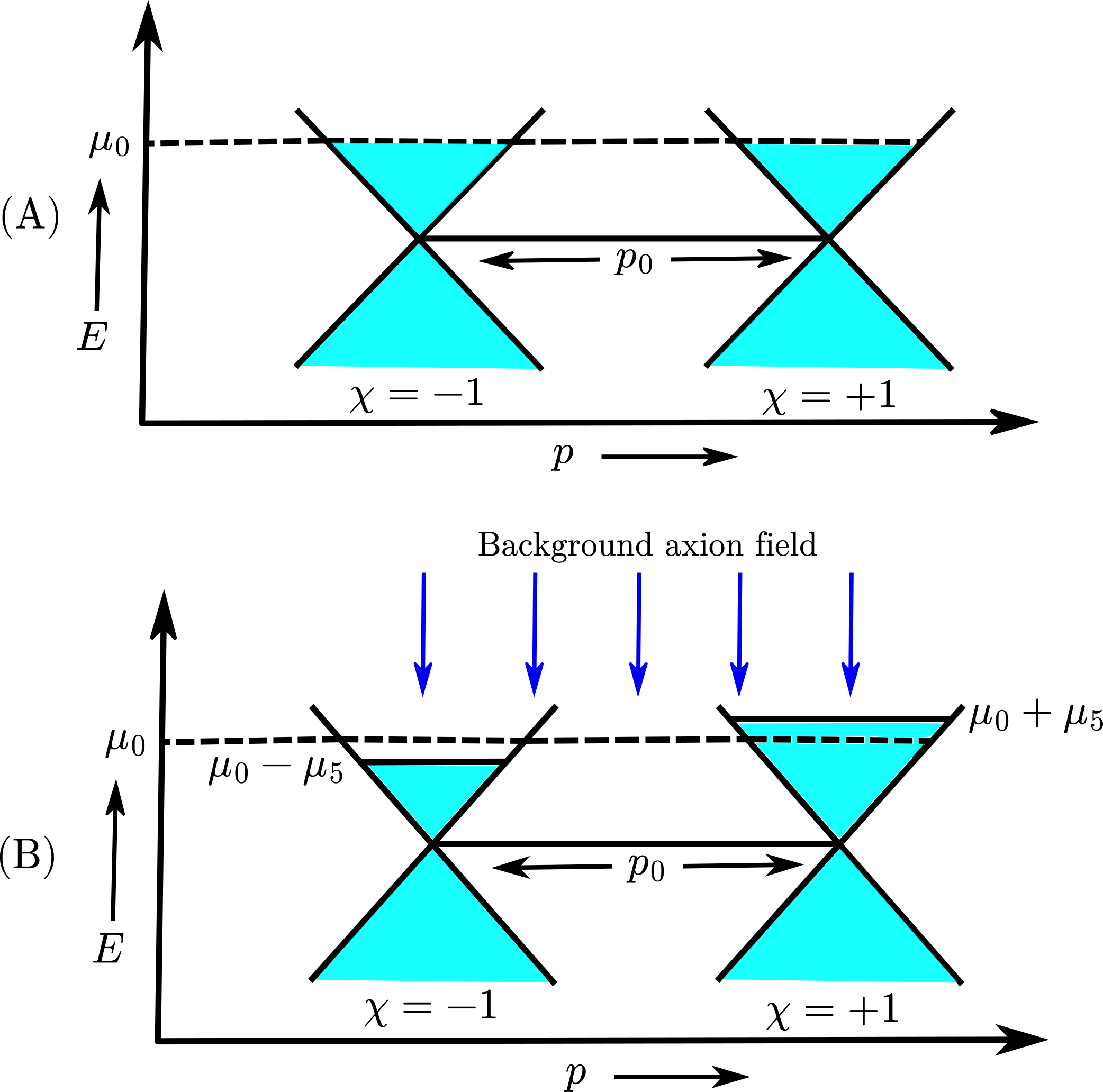}
    \caption{\justifying Schematic band structure near Weyl nodes of opposite chirality in a Weyl semimetal. (A) represents the band structure without the axion background whereas (B) represents the change in band structure in presence of axion field. The equilibrium chemical potential is denoted by $\mu_0 (\mathcal{O}(10^{-3}) \, {\rm eV})$, while the axion-induced axial chemical potential $\mu_5$ shifts the two Weyl nodes in the opposite direction producing a chiral imbalance and consequently a finite CME current. A value of $\mu_5 \sim 10^{-21}$ eV is obtained  for a representative $g_{ae}=10^{-13}.$}
    \label{fig:Weyl_Nodes}
\end{figure}
%
%
\section{Chiral Magnetic Effect in Weyl Semimetals}
\label{sec:chiral_mag}

In the neighborhood of a Weyl node the valence and conduction band touch at a particular point in the momentum space and the semimetal reduces to a two-level system. In such a case, the Hamiltonian can be written as
\begin{equation}\label{eq:H_0}
    H_0 = \chi v_F\,\bm{\sigma}\cdot \mathbf{p} \, ,
\end{equation}
where $v_F$ is the fermi velocity of electrons near the Weyl node, $\sigma_i$ are the Pauli matrices and $\chi=\pm 1$ is the topologically preserved chirality. One can further diagonalise the Hamiltonian in Eq.~\eqref{eq:H_0} to obtain the energy eigenvalues as $\epsilon_\chi =\pm v_F|p|$. 

The Lagrangian for electrons in a Weyl semimetal coupled to an axion field $a(x)$  in the presence of an external magnetic field \textbf{B} is given by,
\begin{equation}\label{eq:axion_lag}
    \mathcal{L} = \bar{\psi}_e i\gamma^\mu D_\mu \psi_e + \bar{\psi}_e \frac{g_{ae}}{2m_e} (\partial_\mu a)\gamma^\mu \gamma^5 \psi_e \, ,
\end{equation}
where $D_\mu = \partial_\mu - i e A_\mu$. The quasi chiral states near the Weyl nodes can be considered effectively massless. For a homogeneous background of ultralight axions, only the time derivative $\dot{a}(t)$ contributes, and the axion-electron interaction reduces to $\mathcal{L}_{ae} = \bar{\psi}_e \left( \mu_5 \gamma^0 \gamma^5 \right) \psi_e$, where the axial chemical potential is defined as,
\begin{equation}\label{eq:chem_pot}
    \mu_5=\frac{g_{ae}}{2m_e}\,\dot a(t) \, .
\end{equation}
Consequently, the axion-electron interaction can be written in terms of expectation of the Hamiltonian as
\begin{equation}\label{eq:del_Hae}
    \langle \delta H_{ae} \rangle = -\langle \psi_e^\dagger \mu_5 \gamma^5 \psi \rangle \, ,
\end{equation}
which leads to an additional energy $\delta H_{ae}=\chi \mu_5$. So the total Hamiltonian near a Weyl node including the axion-electron effect is
\begin{equation}\label{eq:H_tot}
    H=\chi v_F\,\mathbf{\sigma}\cdot \mathbf{p}+\chi \mu_5 \, .
\end{equation}
The axion-electron interaction drives a chiral energy shift $\delta \epsilon_\chi=\chi \mu_5$ as depicted  in Fig. \ref{fig:Weyl_Nodes}. In the presence of an external magnetic field, the  energy  of  the chiral lowest landau level of the quasi charge carriers are given by $\epsilon_{LLL,\chi}=\chi v_F p_z$ and the corresponding group velocity  in the direction of the external magnetic field is given by
\begin{equation}\label{eq:v_LLL}
    v_{LLL,\chi}=\chi v_F
\end{equation} 
Noting that density of states of the lowest landau level is expressed as
\begin{equation}\label{eq:D_LLL}
    D_{LLL}= \frac{eB}{2\pi}
\end{equation} 
The CME current density is given by $J_{CME} \sim (\text{charge}) \times (\text{density of states}) \times (\text{velocity}) \times (\text{occupation change})$:
\begin{equation}\label{eq:JCME_analytic}
  J_{\rm CME} = e \left( \frac{eB}{2\pi} \right) \int \frac{dp_z}{2\pi} \, \chi v_F \, \delta f \, ,
\end{equation}
where $\frac{dp_z}{2\pi}$ is the density of momentum states per unit length and $\delta f$ is the occupation number change due to axion effect. The change in occupation at zero temperature can be approximated as  $\delta f=\delta(\epsilon - \epsilon_F)\,(\chi \,\mu_5),$ 
using the relation $dp_z=d\epsilon/|v_F \chi|$, Eq.~\eqref{eq:JCME_analytic} can be rewritten as
\begin{equation}\label{eq:JCME_analytic_2}
    J_{\rm CME} = e \left( \frac{eB}{2\pi} \right) \int \frac{d\epsilon}{2\pi v_F} \, \chi v_F \, \delta(\epsilon - \epsilon_F)\,(\chi \,\mu_5) \, ,
\end{equation}
Note that, the cancellation of the Fermi velocity can be traced back to the topologically protection of chiral anomaly in Weyl semimetals\,\cite{Burkov:2014aba}. Assuming a single pair of Weyl nodes, the bulk CME current density arising from the two nodes can be written as
\begin{equation}\label{eq:JCME_final_analytic}
    \mathbf{J}_{\rm CME} = \frac{e^2}{2\pi^2} \mu_5\mathbf{B} \, .
\end{equation}
A complementary field-theoretic derivation of Eq.~\eqref{eq:JCME_final_analytic}, based on the Fujikawa method applied to the axion-electron coupling in Eq.~\eqref{eq:axion_lag}, is presented in Appendix~\ref{app:qft_fuji_der}. 

We model the classical axion DM field as $a(t)=a_0\cos(m_a t)$, with local energy density $\rho_{\rm DM}=\frac{1}{2}m_a^2 a_0^2\simeq 0.3\,{\rm GeV/cm^3},$ this implies $\dot a_{\rm amp}=\sqrt{2\rho_{\rm DM}}.$ For a sample of cross-sectional area $A$, if $N_W$ is the number of Weyl nodes pairs in the Brillouin zone,  the resulting bulk CME current driven by the axion background can be written as
\begin{equation}\label{eq:I_CME_tot}
    \mathbf{I}_{\rm CME} = A \, N_W\frac{e^2}{2\pi^2} \mu_5\mathbf{B} =  I_{\rm CME}^{\rm amp} \, \sin(m_a t)  \, ,  
\end{equation}
%
%
\begin{figure}[t]
    \centering
    \includegraphics[width=0.975\linewidth]{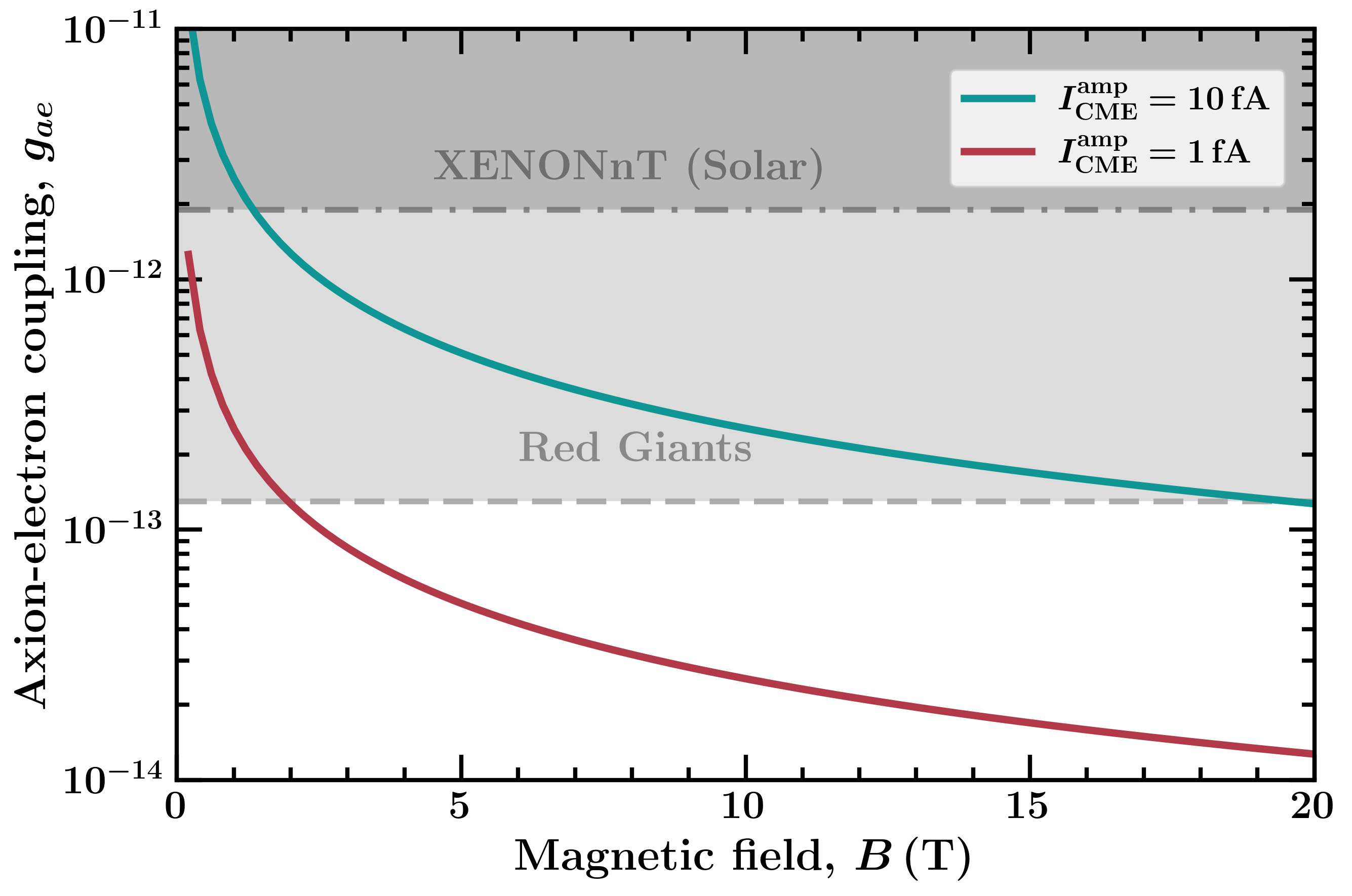}
    \caption{\justifying Projected sensitivity to the axion--electron coupling $g_{ae}$ as a function of magnetic field strength $B$ for a $1~{\rm cm^2}$ chiral semimetal sample. Solid curves correspond to current noise floors of $1 \,{\rm fA}$  and $10 \, {\rm fA}$. The dashed and dotted horizontal lines indicate the representative red-giant energy-loss bound\,\cite{Capozzi:2020cbu} and the direct-detection limit from XENONnT\,\cite{XENON:2022ltv}, respectively, adopted from\,\cite{AxionLimits}.}
    \label{fig:gae_vs_B}
\end{figure}
%
%
where $\mu_5$ is defined in Eq.~\ref{eq:chem_pot} and amplitude of the CME current is defined as $\left| I_{\rm CME}^{\rm amp}\right| = g_{ae} \, N_W \, K_{\rm eff} \, \mathbf{B} \, A$. The specific CME current coefficient $K_{eff}= 3.9\times10^{-3}$ Amp T$^{-1}$ cm$^{-2}$. For a representative setup with $B= 10$ T, a sample area $A=1$ cm$^2$ and the coupling set to the present  astrophysical bound at $g_{ae}=10^{-13}$, the amplitude of the CME current obtained from a single pair of Weyl node is $  \mathbf{I}_{\rm CME}^{\rm amp} \simeq 3.9\times10^{-15}\text{A}.$
This implies that by detecting currents at the femto-ampere (fA) level \cite{Luomahaara:2012squid, Drung:2015ulca, Shingla:2018aadbe1} will enable experiment to probe the unexplored region of axion parameter space as shown in Fig~\ref{fig:gae_vs_B}.
%

%
%
\section{Projected Sensitivity}
\label{sec:experiment_details}

The detection of axion-induced CME currents requires materials that avoid the cancellation of chiral currents. Weyl semimetals like magnetic Kagomes ${\rm Co_3Sn_2S_2}$\,\cite{Liu:2017jxz, Morali:2019phm, Ghimire:2019wzk, Rathod:2020cgs} or Heusler like ${\rm Co_2 Mn Ga}$\,\cite{Belopolski:2020xzr, Kono:2021wfs}, which break time inversion symmetry, have topologically protected Weyl nodes. In these systems, the axion field generates a chiral chemical potential driving a net zero-bias current  that is spin locked in the direction of the external magnetic field. The primary experimental challenge is the detection of currents at the femto-ampere (fA) level which remains within reach of modern SQUID-based and cryogenic current measurement techniques\,\cite{Luomahaara:2012squid, Drung:2015ulca, Shingla:2018aadbe1}.

%
%
\begin{figure}[t]
    \centering
    \includegraphics[width=0.975\linewidth]{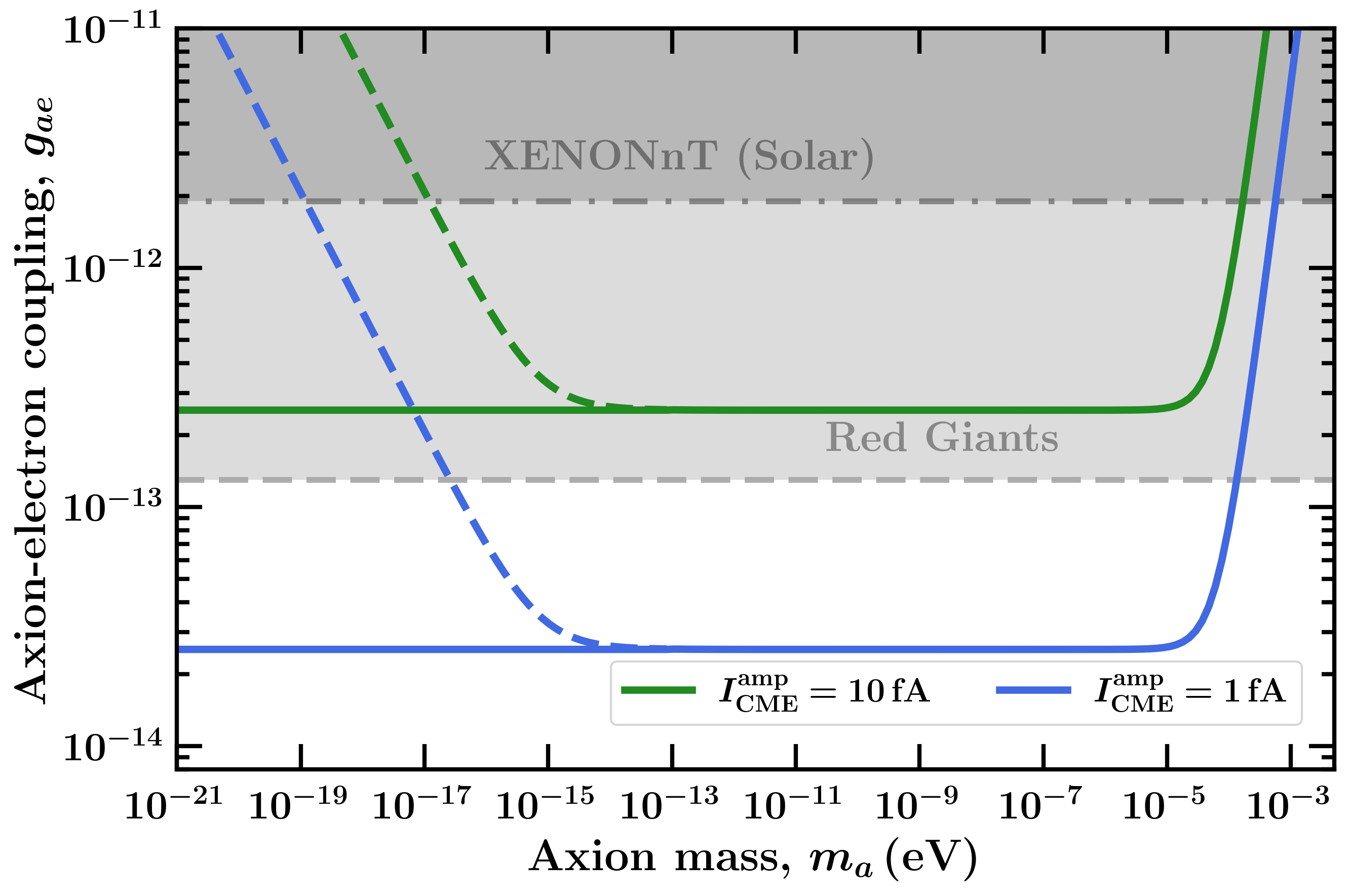}
    \caption{\justifying Projected sensitivity to $g_{ae}$ as a function of axion mass $m_a$ for $B=10$ T for a single Weyl node pair. The dashed curves account for finite carrier relaxation time and frequency dependent SQUID noise and the solid curves represent the sensitivity curves ignoring the measurement noise. The shaded regions indicate the excluded parameter space from stellar cooling\,\cite{Capozzi:2020cbu} and XENONnT\,\cite{XENON:2022ltv}, extracted from\,\cite{AxionLimits}.}
    \label{fig:gae_vs_m_a}
\end{figure}
%
%
While the magnitude of the CME current as defined in Eq.~\ref{eq:I_CME_tot} is independent of axion mass  the frequency  of the generated alternating current scales as  $f \sim m_a,$ this modifies both the material response and the experimental sensitivity of measuring instruments . The shape of the sensitivity curve is governed by two main effects:

(i) \textit{Material Response:} At high frequencies, the CME current is limited by thermal effects due to the  finite carrier relaxation time $\tau$, scaling as $(1+\omega^2\tau^2)^{-1}.$ For $\tau \sim 10^{-11}$ s \,\cite{PhysRevB.111.115309}, this suppression becomes significant  beyond $m_a \gtrsim 10^{-4}$ eV where the sensitivity  gradually drops due to thermal recombination. In such a scenario, the relation between $g_{ae}(m_a)$ and axion mass for a single Weyl node pair can be written as
\begin{equation}
    g_{ae}(m_a)= \frac{\left| \mathbf{I}_{\rm CME}^{\rm amp}\right|}{K_{\rm eff}BA}\left[1+\left(m_a\tau\right)^2\right] \, .
\end{equation}
The  solid lines  in Fig.~\ref{fig:gae_vs_m_a},  represent sensitivity limits assuming  $\mathbf{I}_{\rm CME}$ as 1 fA (blue line) and 10 fA (green line) for our analysis. Expectedly  for $m_a \gtrsim 10^{-4}$ eV, the finite relaxation-time response dominates causing the exclusion curves to rise.

(ii)\textit{Detector Noise:} At low frequencies ($f$),  typically below 1 Hz, the  noise  that scales as $\sim 1/f$ in architectures  like SQUIDs,  degrade sensitivity\,\cite{Clarke:2004squid}. Comparing the CME current   for a single Weyl Node pair with the integrated  SQUID white noise floor curent implies  $\mathbf{I}_{\rm CME}=S_0^{1/2}/\sqrt{T_{eff}}\left(1+\frac{f_k}{m_a}\right)^{1/2}$, where $S_0^{1/2}$ is the SQUID white noise current resolution\,\cite{Gay:2000cca, Shingla:2018aadbe1} and $T_{eff}$ is the integration time. Assuming a  SQUID knee frequency $f_k= \text{1 Hz}$  the corresponding degradation of the  probe sensitivity on the axion parameter space is represented by blue and green dashed line in Figure \ref{fig:gae_vs_m_a}. At frequencies below the knee frequency, the noise dominates which causes the sensitivity curve to rise for low $m_a$. The existing bounds shown in Figs.~\ref{fig:gae_vs_B} and \ref{fig:gae_vs_m_a} are adopted from\,\cite{AxionLimits}. The dashed lines correspond to stellar cooling bounds derived from the observed tip luminosity of the red-giant branch\,\cite{Capozzi:2020cbu}. The dot-dashed line correspond to the XENONnT direct detection bound derived from the absence of excess low-energy electronic recoil events\,\cite{XENON:2022ltv}. In Fig.~\ref{fig:gae_vs_m_a}, we show our limits for $m_a \geq 10^{-21} \, {\rm eV}$, above which ultralight bosonic DM remains consistent with dwarf galaxy observations\,\cite{Zimmermann:2024xvd}. As seen from Fig.~\ref{fig:gae_vs_m_a}, our projected limit obtained for a possible $1 \, {\rm fA}$ current measurement is more stringent than the existing limits for $10^{-17} \, {\rm eV} \lesssim m_a \lesssim 10^{-4} \, {\rm eV}$. The suppression in the projected sensitivity for $m_a \lesssim 10^{-17} \, {\rm eV}$ arises from the frequency-dependent detector noise considered for the SQUID and can be overcome if the detector response does not weaken at lower frequencies. For higher masses, $m_a \gtrsim 10^{-4} \, {\rm eV}$, the limits are suppressed due to thermal effects.

%
\section{Future Experimental Possibilities}
\label{sec:future_prospects}

Sensitivity can be improved by increasing the effective area  $(N_W A)$ of the Weyl semimetal sample by stacking semimetal layers\,\cite{Burkov:2011orp} or by constructing  an array\,\cite{Song:2022nanoph, Song:2023xxy, Kim:2025advs} that increases the total generated CME current given by  Eq.~\ref{eq:I_CME_tot}, thereby allowing the experiment to probe lower values of  $g_{ae}$.

To boost the signal magnitude above the detector noise floor, several approaches may be employed: (i) \textit{Resonant Enhancement:} Coupling the sample to a high-(Q) LC circuit, possibly with a tunable capacitance, can resonantly enhance the measured current prior to readout\,\cite{Sikivie:2013laa}. (ii) \textit{Flux Readout:} Embedding the sample in a superconducting multi-turn coil converts the oscillatory current into a magnetic flux, ($\Phi \sim nLI$), where $n$ denotes the number of turns. The resulting  magnified flux can be coupled to a SQUID, which is particularly sensitive to flux measurements, with noise floors $S_\Phi^{1/2} \sim 10^{-6}\Phi_0/\sqrt{\mathrm{Hz}}$\,\cite{Kahn:2016aff}.
%

%
%
\section{Conclusion}
\label{sec:conclusion}

We have studied the suitability of axion DM detection using the chiral magnetic effect in zero-bias Weyl semimetals in an external magnetic field. We find that axion-electron couplings consistent with existing astrophysical constraints induce oscillatory CME currents at the femto-ampere scale for realistic laboratory parameters. Such signals are within the reach of optimized SQUID-based current measurement schemes, providing a novel laboratory probe of ALPs. For these experiments employing Weyl semimetals as axionic transducers, a current measurement of $1 \, {\rm fA}$ with a $10 \, {\rm T}$  external magnetic field can probe axion-electron coupling $g_{ae} \sim \mathcal{O}(10^{-14})$ for axion mass in the range $10^{-17} \, {\rm eV} \lesssim m_a \lesssim 10^{-4} \, {\rm eV}$, which is about an order of magnitude below the current bounds.
%

%
%
\vspace*{0.4in}
\emph{Note Added:} While this work was in preparation, Ref.\,\cite{Hong:2026jgu} appeared on arXiv with a related study of axion-induced CME current in dense fermionic matter. In contrast to the velocity-suppressed result derived there, we find that for Weyl semimetals having a linear dispersion relation, the topologically protected CME current density emerging from an underlying chiral anomaly retains its universal form, independent of $v_F,$ which is consistent with\,\cite{Chen:2013iga}.

%
%
\section*{Acknowledgments}

We sincerely thank Debraj Choudhury, Arghya Taraphder for their helpful discussions and insightful suggestions. D.B. acknowledges the Council of Scientific and Industrial Research (CSIR), Government of India, for supporting his research under the Research Associateship program through grant no.\,\,09/0079(24106)/2025-EMR-I. P.C. acknowledges MHRD, Government of India for fellowship. T. S. R. acknowledges the Department of Science and Technology, Government of India, under the ANRF Grant Agreement No. MTR/2023/000469 (MATRICS) for financial assistance.
%
%
\appendix
%
%
\section{Anomaly-induced CME current from the Fujikawa method}
\label{app:qft_fuji_der}

In this appendix, we present a complementary derivation of the bulk CME current density of Eq.~\eqref{eq:JCME_final_analytic} using the path-integral formalism. The derivation follows the standard Fujikawa treatment of the axial anomaly, adapted to the case of an axion-electron coupling.

The medium effect enters the Lagrangian  through the  Fermi velocity  scaling of the electron 3-momentum. This can be captured  by an anisotropic kinetic term given by $\mathcal{L} \supset \bar{\psi}_e(i\gamma^0 D_0 + i v_F \gamma^i D_i)\psi_e$. A rescaling of the form $x'^i = x^i/v_F$, $A'_i = v_F A_i$, and $\psi'_e = v_F^{3/2} \psi_e$ leaves the action  invariant  while introducing a non-canonical gauge kinetic term. Thus the topologically protected, chiral anomaly arising  exclusively from the Jacobian of the fermionic path integral measure  remains independent of the Fermi velocity consistent with\,\cite{Chen:2013iga} and can be derived from the following action
\begin{equation}\label{eq:axion_action}
    S_{\psi e} = \int d^4 x \, \left( \bar{\psi}_e i\gamma^\mu D_\mu \psi_e + \bar{\psi}_e \frac{g_{ae}}{2m_e} (\partial_\mu a) \gamma^\mu \gamma^5 \psi_e \right) \, .
\end{equation}
The axion-electron interaction term can be absorbed into a chiral redefinition of  the fermion fields $\psi_e \rightarrow \psi_e'=e^{i\alpha(x)\gamma^5}\psi$, where $\alpha(x)=-\,\frac{g_{ae}}{2m_e}\,a(x).$ This chiral symmetry is  anomalous and the usual Fujikawa formalism yields an anomaly action given by,
\begin{equation}\label{eq:S_anom}
    S_{\rm an} = - \int d^4x \, \frac{e^2}{8\pi^2}\, \alpha(x) \, F_{\mu\nu}\tilde F^{\mu\nu} \, ,
\end{equation}
where $F_{\mu\nu}$ is the electro-magnetic field strength tensor. The anomaly current density associated with this chiral rotation is derived as  $J_\mu(x)={\delta S_{\rm an}}/{\delta A_\mu(x)}$, where the non-vanishing terms are proportional to  $\dot a \mathbf{B}$, where $\mathbf{B}$ represent the external applied magnetic field. The resulting anomaly current density associated with  this  chiral rotation is given by

\begin{equation}\label{eq:J_CME_qft}
   \mathbf{J}_{\rm CME} =\frac{e^2}{2\pi^2} \mu_5\mathbf{B} \, ,
\end{equation}
which can be identified with the bulk CME current in the Weyl semimetals generated in a single pair of Weyl nodes and matches with the result obtained in Eq.~\eqref{eq:JCME_final_analytic}.
%

\bibliographystyle{JHEP}
\bibliography{refs}

\end{document}